\documentclass[12pt,preprint]{emulateapj}
\usepackage{apjfonts}

\newcommand{\reff}{r_\mathrm{eff}}
\newcommand{\ger}{K2000 }
\newcommand{\rh}{r_h}
\newcommand{\vh}{v_h}
\newcommand{\rhoh}{\rho_h}
\newcommand{\kpc}{\mathrm{kpc}}
\newcommand{\pc}{\mathrm{pc}}
\newcommand{\Mpc}{\mathrm{Mpc}}
\newcommand{\Gyr}{\mathrm{Gyr}}
\newcommand{\kms}{\mathrm{km \, s^{-1}}}
\newcommand{\rhodm}{\rho_\mathrm{DM}}
\newcommand{\zform}{z_\mathrm{form}}

\newcommand{\zssp}{z(\tau_0)}
\newcommand{\dbar}{\delta_\mathrm{bar}}
\newcommand{\dhalo}{\delta_\mathrm{halo}}
\newcommand{\dobs}{\delta_\mathrm{obs}}

\newcommand{\lsun}{L_\odot}
\newcommand{\msun}{M_\odot}
\newcommand{\mstar}{M_\ast}
\newcommand{\disk}{^{\mathrm{D}}}
\newcommand{\spir}{^{\mathrm{S}}}
\newcommand{\el}{^{\mathrm{E}}}
\newcommand{\rhoavdm}{\left< \rho_\mathrm{DM}\right>}

\shorttitle{The assembly epoch of Coma early-type galaxies}
\shortauthors{J. Thomas et al.}

\begin{document}

\title{Dark matter scaling relations and the assembly epoch of Coma early-type galaxies}

\author{J. Thomas\altaffilmark{1}, R. P. Saglia, R. Bender}
\affil{Universit\"atssternwarte M\"unchen, Scheinerstra\ss e 1, D-81679 M\"unchen, Germany\\
Max-Planck-Institut f\"ur Extraterrestrische Physik, Giessenbachstra\ss e, D-85748 Garching, Germany}
\altaffiltext{1}{E-mail: jthomas@mpe.mpg.de}
\author{D. Thomas}
\affil{Institute of Cosmology and Gravitation, Mercantile House, University of Portsmouth, Portsmouth, PO1 2EG, UK}
\author{K. Gebhardt}
\affil{Department of Astronomy, University of Texas at Austin, C1400, Austin, TX78712, USA}
\author{J. Magorrian} 
\affil{Theoretical Physics, Department of Physics, University of Oxford, 1 Keble Road, Oxford U.K., OX1 3NP}
\author{E.~M. Corsini}
\affil{Dipartimento di Astronomia, Universit\`a di Padova, vicolo dell'Osservatorio 3, I-35122 Padova, Italy}
\and
\author{G. Wegner}
\affil{Department of Physics and Astronomy, 6127 Wilder Laboratory, Dartmouth College, Hanover, NH 03755-3528, USA}

\begin{abstract}
Axisymmetric, orbit-based dynamical models are used to derive dark matter scaling
relations for Coma early-type galaxies. From faint to bright galaxies halo core-radii
and asymptotic circular velocities increase. Compared to spirals of the same
brightness, the majority of Coma early-types -- those with old stellar populations --
have similar halo core-radii but more than 2 times larger asymptotic halo velocities. The average dark matter density inside 
$2 \, \reff$ decreases with increasing luminosity and is $6.8$ times larger 
than in disk galaxies of the same $B$-band luminosity. 
Compared at the same stellar mass, dark matter densities in ellipticals are $13.5$ times 
higher than in spirals. Different baryon concentrations in
ellipticals and spirals cannot explain the higher dark matter density
in ellipticals. Instead, the assembly redshift (1+z) of Coma early-type halos is likely 
about two times larger than of comparably bright spirals. Assuming that local spirals typically
assemble at a redshift of one, the majority of bright Coma early-type galaxy halos 
must have formed around $z \approx 2-3$. For about half of our Coma galaxies the 
assembly redshifts match with constraints derived from stellar populations.
We find dark matter densities and
estimated assembly redshifts of our observed Coma galaxies in reasonable agreement with 
recent semi-analytic galaxy formation models. 
\end{abstract}


\section{Introduction}
\label{sec:intro}
Present-day elliptical galaxies are known to host mostly old stellar populations 
(\citealt{Tra00}, \citealt{Ter02}, \citealt{DTho05}).
Whether their stars have formed in situ or whether ellipticals assembled
their present-day morphology only over time (for example by mergers) is less clear.
An important clue on the assembly redshift of a galaxy is provided by its
dark matter density. For example, in the simple spherical collapse model \citep{Gun72} 
the average density 
of virialized halos is proportional to the mean density of the universe at the
formation epoch: halos which form earlier become denser. Similarly, in cosmological
$N$-body simulations, the concentration 
(and, thus, the inner density) is found to be higher in 
halos that have assembled earlier (e.g. \citealt{Nav96}, \citealt{Wec02}). In addition to this connection between formation epoch
and halo density, the final halo mass distribution also depends on the interplay 
between dark matter and baryons during the actual galaxy formation process 
(e.g. \citealt{Blu86}, \citealt{Bin01}). Then, 
the properties of galaxy halos provide valuable information about when and how a 
galaxy has assembled its baryons.

Despite its cosmological relevance, the radial distribution of dark (and luminous) 
mass in early-type galaxies is not well known: because
of the lack of cold gas as a dynamical tracer, masses are difficult to
determine. Stellar dynamical models require the
exploration of a galaxy's orbital structure and have only recently become available
for axisymmetric or more general systems 
(\citealt{Cre99}, \citealt{Geb00}, \citealt{Tho04}, \citealt{Val04}, \citealt{Cap05},
\citealt{deLor07}, \citealt{Cha08}, \citealt{vdB08}). 
Scaling relations for the inner dark matter
distribution in early-types have by now only
been reported for round and non-rotating galaxies (\citealt{Kr00}, \citealt{G01})
and spirals (\citealt{PSS1,PSS2} and \citealt{Kor04}). The aim 
of the present paper is to provide empirical scaling relations for generic cluster early-types
(flattened, with different degrees of rotation). In particular, this paper is
focussed on the inner dark matter density and its implications on the assembly redshift
of elliptical galaxy halos.

The paper is organized as follows.
In Sec.~\ref{sec:data}, we review the galaxy sample and its modelling. 
Dark matter scaling relations are presented in Sec.~\ref{sec:scaling}. Sec.~\ref{sec:density} 
is dedicated to the dark matter density. The effect of baryons on the dark matter
density is discussed in Sec.~\ref{sec:contraction}, while Sec.~\ref{sec:assembly} deals with the 
halo assembly redshift. In Sec.~\ref{sec:semi-analytic} our results are compared to 
semi-analytic galaxy formation models. A summary is given in Sec.~\ref{sec:summary}.
In the following, we assume that the Coma cluster is at
a distance of $d = 100 \, \Mpc$.

\section{Galaxy sample, models and basic definitions}
\label{sec:data}
The dark halo parameters discussed in this paper are derived from the axisymmetric,
orbit-based dynamical models of 
bright Coma galaxies presented in \citet{Tho07}. The original sample comprises two cD galaxies, nine
ordinary giant ellipticals and six lenticular/intermediate type galaxies with
luminosities between $M_B = -18.79$ and $M_B = - 22.56$. The spectroscopic and photometric 
observations are discussed in \citet{Jor96}, \citet{Meh00}, \citet{Weg02} and \citet{Cor08}.
Our implementation of Schwarzschild's (1979) orbit superposition technique for axisymmetric
potentials is described in \citet{Tho04,Tho05}. For a detailed discussion of all the
galaxy models the reader is referred to \citet{Tho07}.

Three of the 17 galaxies from \citet{Tho07} are excluded from the analysis below.
Firstly, we do not consider the two central cD galaxies (GMP2921 and GMP3329; GMP
numbers from \citealt{GMP}), because their dark matter profiles may be affected by the 
cluster halo. Secondly, we omit the E/S0 galaxy GMP1990, whose 
mass-to-light ratio is constant out to $3 \, \reff$. The galaxy either has no dark matter 
within this radius, or its dark matter density follows closer the stellar light 
profile than in any other Coma galaxy. In either case, the mass structure of this
object is distinct from the rest of the sample galaxies. 
In addition to the remaining 14 galaxies we consider
two further Coma galaxies for 
which we collected data recently (GMP3414, GMP4822). 
The models of these galaxies are summarized in
App.~\ref{app:3414}.  

Similar dynamical models as used here have been applied to the inner regions of ellipticals,
where it has been assumed that mass follows light (e.g. \citealt{Geb03}, 
\citealt{Cap05}). In contrast, our models explicitly include
a dark matter component (cf. \citealt{Tho07}). We probed for two parametric profiles.
Firstly, logarithmic halos
\begin{equation}
\label{logdef}
\rho_\mathrm{DM}(r) = \frac{\vh^2}{4 \pi G} \frac{3 \rh^2+r^2}{(\rh^2+r^2)^2},
\end{equation}
which posses a constant-density core of size $\rh$ and have an asymptotically constant 
circular velocity $\vh$. The central density of these halos reads
\begin{equation}
\label{logrho0def}
\rhoh = \frac{3 \vh^2}{4 \pi G \rh^2}.
\end{equation}
Secondly, NFW-profiles
\begin{equation}
\label{nfwdef}
\rho_\mathrm{DM}(r) \propto \frac{1}{r(r+r_s)^2},
\end{equation}
which are found in cosmological $N$-body simulations \citep{Nav96}. 
The majority of Coma 
galaxies are better fit with logarithmic halos, but the significance over NFW halo profiles
is marginal. Even if the best fit is obtained with an NFW-halo, then the inner regions
are still dominated by stellar mass (cf. \citealt{Tho07}). In this sense,
our models maximize the (inner) stellar mass.

In Sec.~\ref{sec:scaling} we will only discuss results based on logarithmic halo 
fits (i.e. we use the halo parameters from columns 5 and 6 in Tab.~2 of \citealt{Tho07} and 
columns 3 and 4 in Tab.~\ref{tab:3414} of App.~\ref{app:3414}, respectively). 
While these are not necessarily the more realistic profiles, they 
minimize systematics in the comparison with published scaling relations
for spirals that were performed using cored profiles similar to our 
logarithmic halos. The NFW fits are used in Sec.~\ref{sec:contraction}. 

The $B$-band luminosities of Coma galaxies used in this paper are taken from 
Hyperleda. We adopt a standard uncertainty of $\Delta M_B = 0.3$ to
account for zero-point uncertainties, systematic errors in the sky subtraction,
seeing convolution, profile extrapolation and others. Effective radii are taken from
\citet{Jor95} and \citet{Meh00}. Here we estimate the errors to be $\Delta \log \reff = 0.1$.
This is slightly higher than the uncertainties given in \citet{Jor95}, but accounts
for possible systematic errors \citep{Sag97}.
Stellar masses were computed from our best-fit stellar mass-to-light ratios $\Upsilon$ 
and $R$-band luminosities of \citet{Meh00}.
In case the best fit is obtained with a logarithmic halo, $\Upsilon$ is
taken from column 4 of Tab.~2 in \citet{Tho07}. In case of
an NFW fit, $\Upsilon$ comes from column 8 of the same table. The best-fit stellar mass-to-light
ratios of GMP3414 and GMP4822 are given in column 1 of Tab.~\ref{tab:3414}
(cf. App.~\ref{app:3414}). 

\begin{deluxetable}{lcccc}
\tablecolumns{5}
\tabletypesize{\small}
\tablewidth{0pt}
\tablecaption{Galaxy parameters shown in Fig.~\ref{fig:rhvh}\label{tab:gals1}}
\tablehead{
\colhead{galaxy} &
\colhead{$\log \frac{L_B}{\lsun}$}&
\colhead{$\log \frac{\mstar}{\msun}$}&
\colhead{$\log \frac{\rh}{\kpc}$}&
\colhead{$\log \frac{\vh}{\kms}$}\\
\colhead{(1)} &
\colhead{(2)}&
\colhead{(3)}&
\colhead{(4)}&
\colhead{(5)}
}
\startdata
0144 & $10.61$ & $11.56 \pm 0.12$ & $ 0.64 \pm  0.31 $ & $ 2.33 \pm  0.10$\\
0282 & $10.46$ & $11.60 \pm 0.12$ & $ 1.23 \pm  0.24 $ & $ 2.70 \pm  0.10$\\
0756 & $10.89$ & $11.13 \pm 0.12$ & $ 1.10 \pm  0.09 $ & $ 2.33 \pm  0.10$\\
1176 & $10.31$ & $10.73 \pm 0.13$ & $ 0.53 \pm  0.18 $ & $ 2.30 \pm  0.11$\\
1750 & $10.75$ & $11.58 \pm 0.12$ & $ 1.27 \pm  0.95 $ & $ 2.70 \pm  0.23$\\
2417 & $10.60$ & $11.43 \pm 0.12$ & $ 1.38 \pm  0.59 $ & $ 2.70 \pm  0.38$\\
2440 & $10.30$ & $11.23 \pm 0.12$ & $ 1.04 \pm  0.15 $ & $ 2.68 \pm  0.13$\\
3414 & $10.13$ & $11.02 \pm 0.15$ & $ 0.99 \pm  0.54 $ & $ 2.55 \pm  0.26$\\
3510 & $10.34$ & $11.28 \pm 0.13$ & $ 1.07 \pm  0.39 $ & $ 2.46 \pm  0.21$\\
3792 & $10.58$ & $11.56 \pm 0.13$ & $ 1.18 \pm  0.35 $ & $ 2.74 \pm  0.22$\\
3958 & $ 9.70$ & $10.81 \pm 0.13$ & $ 0.83 \pm  0.35 $ & $ 2.44 \pm  0.29$\\ 
4822 & $10.70$ & $11.69 \pm 0.16$ & $ 1.11 \pm  0.50 $ & $ 2.74 \pm  0.37$\\
4928 & $11.08$ & $12.06 \pm 0.14$ & $ 1.46 \pm  0.39 $ & $ 2.71 \pm  0.19$\\
5279 & $10.72$ & $11.59 \pm 0.12$ & $ 1.45 \pm  0.48 $ & $ 2.68 \pm  0.28$\\
5568 & $10.79$ & $11.89 \pm 0.12$ & $ 1.82 \pm  0.39 $ & $ 2.81 \pm  0.20$\\
5975 & $10.47$ & $11.04 \pm 0.12$ & $ 0.23 \pm  0.30 $ & $ 2.30 \pm  0.11$
\enddata
\tablecomments{(1) Galaxy id from \citet{GMP}; (2) galaxy $B$-band luminosity $L_B$; (3) stellar mass $\mstar$ in solar 
units; (4,5) logarithmic-halo core-radius $\log \rh/\kpc$ and
circular velocity $\log \vh/\kms$.}
\end{deluxetable}

\section{Dark matter scaling relations}
\label{sec:scaling}
\begin{figure*}\centering
\begin{minipage}{140mm}
\centering
\includegraphics[width=135mm,angle=0]{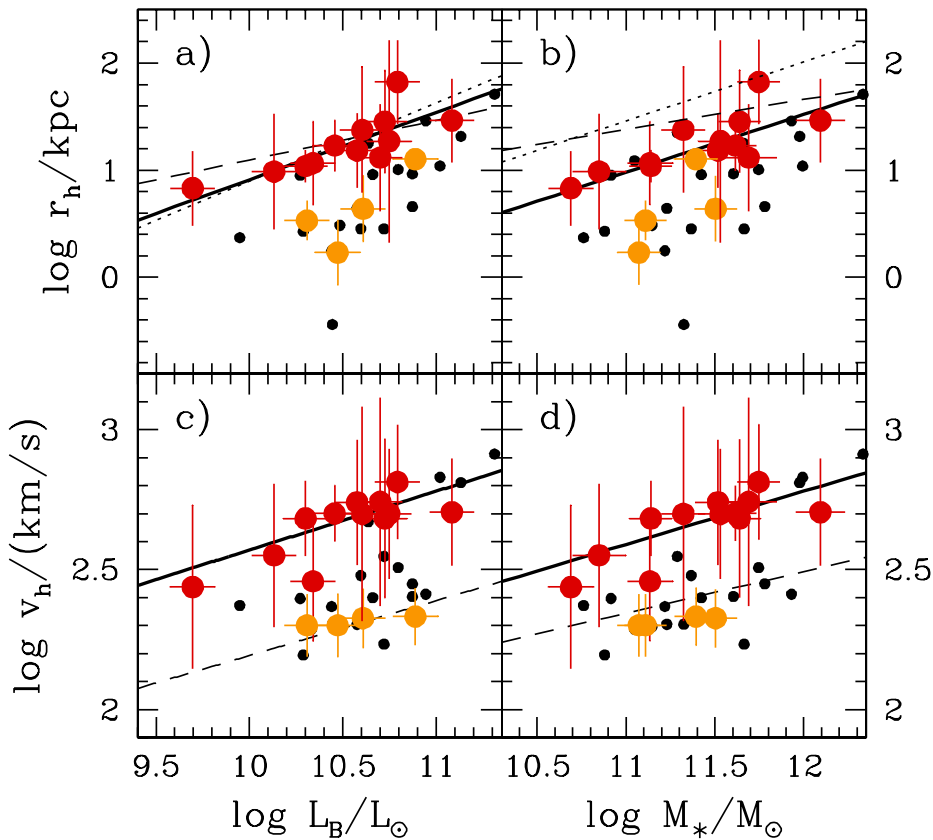}
\caption[]
{Halo core-radius $\rh$ and circular velocity $\vh$ versus $B$-band luminosity $L_B$ (a,c) 
and versus stellar mass $\mstar$ (b,d). Large symbols: Coma ellipticals (red: central stellar population age $\tau_0 > 6 \, \Gyr$, 
orange: $\tau_0 < 6 \, \Gyr$, details in the text); thick solid lines: fits to galaxies
with $\tau_0 > 6 \, \Gyr$; small symbols: round early-types from \citet{G01}; 
dotted: spiral galaxy scaling relations from \citet{PSS1,PSS2}; short-dashed: 
spiral galaxy scaling relations from \citet{Kor04}.}
\label{fig:rhvh}
\end{minipage}                  
\end{figure*}   

Fig.~\ref{fig:rhvh} shows the scalings of dark halo core-radii $\rh$ and halo asymptotic
circular velocities $\vh$ with $B$-band luminosity $L_B$ and stellar mass $\mstar$ (the
corresponding galaxy parameters with errors are listed in Tab.~\ref{tab:gals1}). 
Both, halo core-sizes and halo circular velocities tend to increase with luminosity and
mass. The case for a correlation between $\vh$ and $L_B$ is weak, if the 
sample as a whole is considered (cf. column 8 of Tab.~\ref{tab:fits}).
However, four galaxies (GMP0144, GMP0756, GMP1176 and GMP5975) separate from the rest of the
sample galaxies in having both noticeably smaller $\rh$ and $\vh$. These galaxies
are shown in light color in Fig.~\ref{fig:rhvh}. As a general trend, halo parameters
tend to scale more tightly with luminosity and mass when these galaxies are omitted.
The solid lines in Fig.~\ref{fig:rhvh} show corresponding log-linear fits\footnote{Fits for 
this paper are performed with the routine fitexy of 
\citet{press}.}. For comparison, in Tab.~\ref{tab:fits} we give both, fits to all Coma 
galaxies as well as fits to the subsample without the four galaxies offset in Fig.~\ref{fig:rhvh}.
The difference between these four galaxies and the rest of the sample is further discussed
below.

\begin{figure}\centering
\plotone{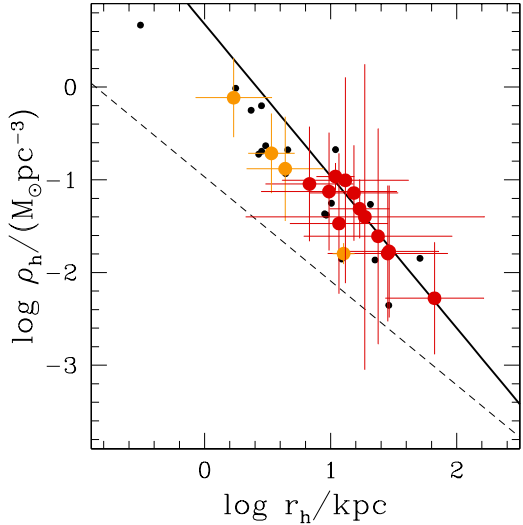}
\caption[Central dark matter scaling relations]
{Central dark matter density $\rhoh$ versus 
halo core-radius $\rh$. Symbols and lines as in Fig.~\ref{fig:rhvh}.}
\label{fig:rhoh_rh}
\end{figure}

The logarithmic halos of equation (\ref{logdef}) have two free parameters. Any and
pair of $\rh$, $\vh$ or $\rhoh$ characterizes a specific halo.
Fig.~\ref{fig:rhoh_rh} shows a plot of $\rhoh$ versus $\rh$. Both halo
parameters are clearly correlated. A linear relation fits the points with a minimum $\chi^2_\mathrm{red} = 0.41$ 
(per degree of freedom; cf. Tab.~\ref{tab:fits}). This rather low value partly derives
from a degeneracy between the halo parameters in the dynamical modeling 
(e.g. \citealt{Ger98}, \citealt{Tho04}) which correlates the errors in both quantities.
In Tab.~\ref{tab:fits}, such a correlation between the errors is not taken
into account and the $\chi^2_\mathrm{red}$ might be thus underestimated. A 
$\chi^2_\mathrm{red}$ much larger than unity would indicate some intrinsic
scatter in Fig.~\ref{fig:rhoh_rh}, whereas the 
low $\chi^2_\mathrm{red}$ quoted in Tab.~\ref{tab:fits} formally rules out any 
intrinsic scatter. 
Note that dark matter halos in cosmological $N$-body simulations can be approximated by
a two-parameter family of halo models, where the parameters are correlated qualitatively in
a similar way as revealed by Fig.~\ref{fig:rhoh_rh} (e.g. \citealt{Nav96}, \citealt{Wec02}),
but with some intrinsic scatter.

The four galaxies offset in Fig.~\ref{fig:rhvh} are also slightly offset
in Fig.~\ref{fig:rhoh_rh}. However, given the large uncertainties, this is not
significant and Fig.~\ref{fig:rhoh_rh} is consistent with the halos of the four galaxies
belonging to the same one-parameter family as established by the remaining Coma 
galaxies. This implies that the four galaxies primarily differ in the amount of stellar light 
(and stellar mass, respectively) that is associated to a given halo. Noteworthy, the four
galaxies have stellar ages $\tau_0 < 6 \, \Gyr$ \citep{Meh03}, while
all other Coma galaxies are significantly older (mostly $\tau_0 \ga 10 \, \Gyr$).
A mere stellar population effect, however, is unlikely to explain the offset of the four 
galaxies. In this case differences to other Coma galaxies should vanish when galaxies
are compared at the same stellar mass, which is not consistent with Fig.~\ref{fig:rhvh}b/d. 
It should be noted, though, that the stellar masses used here are taken from the
dynamical models. A detailed comparison with mass-to-light ratios from stellar population
synthesis models is planed for a future publication (Thomas et al., in preparation).

In Fig.~\ref{fig:rh_reff}a we plot $\rh$ 
against $\reff$. Larger core-radii are found in more extended galaxies.
Three of the four galaxies with young cores are
again offset. These three galaxies have similar
$\reff$ than other Coma galaxies of the same luminosity (cf. Fig.~\ref{fig:rh_reff}b). 
This makes a higher baryon concentration unlikely to be the cause of their small halo 
core-radii. An exceptional case is GMP0756: it has a small core-radius,
a ratio $\rh/\reff$ which is typical for Coma galaxies with old stellar populations
and a relatively small $\reff$. 
The scalings of $\reff$ and $\rh$ with luminosity
in Coma galaxies with old stellar populations imply a roughly constant 
ratio $\rh/\reff \approx 3$. Moreover, disk galaxies of the same luminosity show a
similar ratio.

The fact that all four galaxies offset in Figs.~\ref{fig:rhvh} and \ref{fig:rh_reff} 
appear at projected cluster-centric distances $D > 1 \, \mathrm{Mpc}$ and have
young stellar populations suggests that they may have entered the Coma cluster
only recently. We will come back to this point in the next Sec.~\ref{subsec:round}.

\subsection{Comparison with round and non-rotating early-types}
\label{subsec:round}
\begin{figure}\centering
\plotone{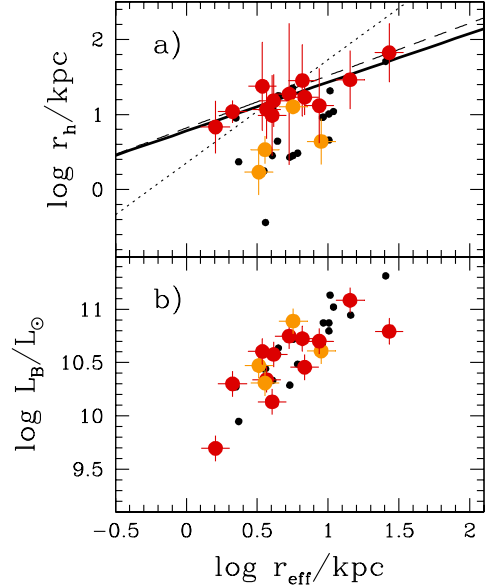}
\caption[]
{Halo core-radius $\rh$ (a) and $B$-band luminosity (b) versus effective radius $\reff$. 
Large symbols: Coma
galaxies; small dots: round galaxies from \citet{G01}. Lines
in panel (a) as in Fig.~\ref{fig:rhvh}.}
\label{fig:rh_reff}
\end{figure}

\citet{Kr00} and \citet{G01} studied dark matter halos of
21 nearly round (E0-E2) and non-rotating galaxies (\ger in the following). \ger galaxies 
have similar luminosities $M_B$ and half-light radii $\reff$ as ours, but 
the \ger sample contains a mixture of field ellipticals and galaxies in the Virgo and 
Fornax clusters. The dynamical models of \citet{Kr00} differ in
some respects from the ones described in Sec.~\ref{sec:data} 
and this will be further discussed below. However, in their mass
decomposition \citet{Kr00} assumed the same halo
profile as in equation (\ref{logdef}), such that we can directly compare 
their halo parameters to ours (cf. small black dots in Figs.~\ref{fig:rhvh} - \ref{fig:rh_reff}).

We find halo parameters of both samples in the same range, 
but Coma galaxies of the same $L_B$ have on average larger 
halo core-radii than \ger galaxies. 
However, the halos of the \ger galaxies themselves are not different from the
ones around Coma early-types, as both belong to the same one-parameter family 
(cf. Fig.~\ref{fig:rhoh_rh}).
The main difference is that \ger galaxies are brighter 
(and have higher stellar mass) than Coma early-types with a similar halo. 
Can this be an artifact related to differences in the dynamical models?

Many of the \ger models are based on $B$-band photometry,
while we used $R_C$-band images for the Coma galaxies. Elliptical galaxies become 
bluer towards the outer parts and $B$-band light profiles are slightly shallower 
than $R$-band profiles. Likewise, mass profiles of galaxies are generally shallower than
their light profiles, such that there might be less need for dark matter in $B$-band
models than in $R$-band models. \citet{Kr00} checked for this by
modelling one galaxy (NGC3379) in both bands and found comparable results. The photometric
data is therefore unlikely to cause the differences between the two samples.

\ger galaxies were modelled assuming spherical symmetry. Not all apparently round galaxies
need to be intrinsically spherical. Neglecting the flattening along the 
line-of-sight can result in an underestimation of a galaxy's mass (e.g. \citealt{Tho07b}).
Based on the average intrinsic flattening of ellipticals in the luminosity 
interval of interest here, \citet{Kr00} estimated that the assumption of spherical symmetry 
should affect mass-to-light ratios only at the 10 percent level. We expect the effect 
on $\vh$ to be correspondingly small. In addition, it is not obvious why spherical symmetry 
should enforce systematically small $\rh$, such that the different symmetry assumptions
are also unlikely to explain the more extended cores and higher circular velocities 
in Coma galaxy halos.

Since the shape of a galaxy is related to its evolutionary history, the round and 
non-rotating \ger galaxies could be intrinsically different from the mostly 
flattened and rotating Coma galaxies. Structural differences could also be related
to the fact that \ger galaxies
are located in a variety of environments, with less galaxies in high
density regions like Coma. For example, stellar population models indicate that field ellipticals are 
on average younger and have more extended star formation histories than cluster 
galaxies \citep{DTho05}. But a mere difference in the stellar populations 
can not explain the difference between \ger and Coma galaxies, as in this
case the scalings with $\mstar$ should be similar in both samples.
This is ruled out by Fig.~\ref{fig:rhvh}b/d.

Noteworthy, most of the \ger galaxies in Figs.~\ref{fig:rhvh} - \ref{fig:rh_reff} appear
similar to the four Coma galaxies with distinctly small $\rh$ and $\vh$. As it has been discussed above, 
these galaxies may have entered the Coma cluster only recently and -- in this respect -- 
are more representative for a field galaxy population rather than being genuine
old cluster galaxies. A consistent explanation for both, the offset between young and
old Coma galaxies on the one side and the difference between old Coma galaxies and 
\ger galaxies on the other would then be that field galaxies have lower $\rh$ and 
lower $\vh$ than cluster galaxies of the same 
stellar mass. Because there are more field galaxies in the
\ger sample than in Coma, Coma galaxy halos would be
expected to have on average larger cores and to be more massive 
(consistent with Fig.~\ref{fig:rhvh}). A 
larger comparison sample of field elliptical halos is required to conclude finally upon this
point.
\begin{figure*}\centering
\begin{minipage}{140mm}
\centering
\includegraphics[width=135mm,angle=0]{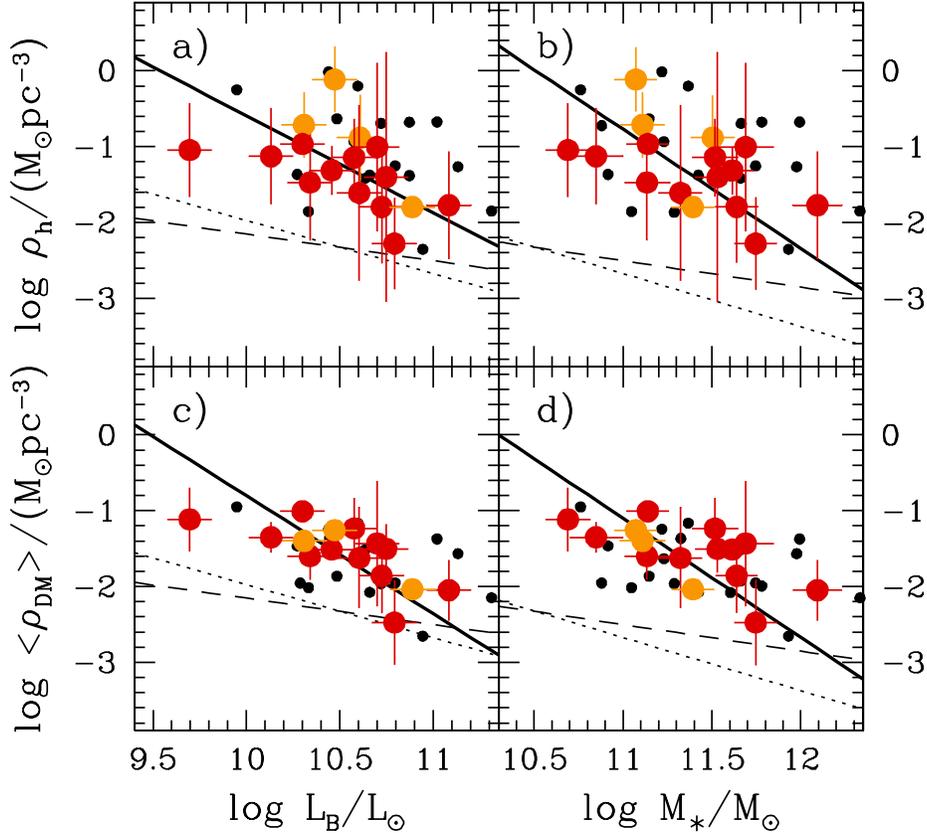}
\caption{Central halo density $\rhoh$ and average dark matter density $\rhoavdm$
inside $2 \, \reff$
versus luminosity $L_B$ (a,c) and stellar
mass $\mstar$ (b,d). Solid lines: fits including all Coma galaxies; 
dotted and dashed lines: spiral galaxy scaling relations 
as in Fig.~\ref{fig:rhvh}.}
\label{fig:rho_l}
\end{minipage}
\end{figure*}

\subsection{Comparison with spiral galaxies}
Two independent derivations of dark matter
scaling relations for spiral galaxies are included in Figs.~\ref{fig:rhvh} -
\ref{fig:rh_reff} through the dotted and dashed lines. Dotted lines show 
scaling relations from \citet{PSS1,PSS2}. They are based on maximum-disk rotation-curve 
decompositions with the halo density from equation (\ref{logdef}). 
\citet{PSS1,PSS2} give halo core-radii scaled by the optical
disk radius. To reconstruct the underlying relationship between $\rh$ and $L_B$, we
follow \citet{G01} and assume exponential disks with
\begin{equation}
\label{spireff}
\left( \frac{\reff\spir}{\kpc} \right) = 8.4 \left( \frac{L_B}{10^{11} L_\odot} \right)^{0.53}
\end{equation}
(the empirical fit in \citealt{G01} has been transformed to our distance scale).

Dashed lines in Figs.~\ref{fig:rhvh} - \ref{fig:rh_reff} fit the combined sample
of 55 rotation curve decompositions of \citet{Kor04}. These authors discuss
various halo profiles, but we here only consider non-singular 
isothermal dark matter halos. Though these are most similar to 
equation (\ref{logdef}), isothermal cores and circular velocities are not exactly identical
as in logarithmic halos. To account for the difference, we fitted a 
logarithmic halo to a non-singular isothermal density profile 
(cf. Tab.~4.1 of \citealt{Bin87}). The fit was restricted to the region with
kinematic data (typically inside two core-radii). We found that the logarithmic halo fit yields 
a 3 percent larger core-radius $\rh$ and a 10 percent smaller circular velocity
$\vh$. The central halo density is reproduced to 0.001 dex (not surprising given that the 
cores of the two profiles were matched). Thus, under 
the assumption that fits performed with the two 
profiles indeed match inside two core-radii, a correction of the derived
halo parameters is not needed. Scaling relations from \citet{Kor04} 
are shown in Figs.~\ref{fig:rhvh} - \ref{fig:rhoh_rh} without any correction.

\citet{PSS1,PSS2} and \citet{Kor04} discuss the scaling of disk galaxy halos with 
$B$-band luminosity. In order to compare
early-types and spirals also at the same stellar masses, we used 
\begin{equation}
\label{eq:mlspir}
\left( \frac{M_\ast}{L_B} \right)\spir = 2 \times \left( \frac{L_B}{10^{11} \, L_\odot} \right)^{0.33}
\end{equation}
for the stellar mass-to-light ratios of disks. Equation (\ref{eq:mlspir}) is
derived from the Tully-Fisher and stellar-mass Tully-Fisher relations of \citet{Bel01}.

Both in luminosity and in stellar mass, spirals and ellipticals follow similar
global trends. However, while old Coma early-types have halo core-radii of similar
size as spirals with the same $B$-band luminosity, the asymptotic halo circular velocities 
are $2.4$ times higher than in corresponding spirals. In contrast, early-types with young
central stellar populations have about 4 times smaller core-radii than spirals, but similar
asymptotic halo velocities. When galaxies are compared at the same stellar
mass, then differences between ellipticals and
spirals become larger (60 percent smaller $\rh$ and $1.8$ times higher $\vh$
in old Coma early-types compared to spirals; 90 percent smaller $\rh$ and 20 percent 
smaller $\vh$ in Coma galaxies with young central stellar populations).
In addition, the halos of early-types and spirals do not belong
to the same one-parameter family (cf. Fig.~\ref{fig:rhoh_rh}). At a given $\rh$
dark matter densities in ellipticals are about $0.5$ dex higher than in spirals.

\section{The dark matter density}
\label{sec:density}
Fig.~\ref{fig:rho_l} shows scaling laws for dark matter densities. 
The central dark matter density $\rhoh$ (cf. equation \ref{logrho0def}) 
of the logarithmic halo 
fits is plotted in panels (a) and (b) versus luminosity and stellar mass. 
For panels (c) and (d), the best-fit dark matter halo of each galaxy (being either
logarithmic or NFW) is averaged within $2 \, \reff$:
\begin{equation}
\label{avrhodef}
\rhoavdm \equiv \frac{3}{4 \pi} \, \frac{M_\mathrm{DM} (2 \, \reff)}{(2 \, \reff)^3}
\end{equation}
Here, $M_\mathrm{DM}(r)$ equals the cumulative dark mass inside a sphere with radius $r$.
Average dark matter densities are quoted in column (2) of Tab.~\ref{tab:gals2}.

The general trend for both densities is to decrease with increasing galaxy luminosity.
Thereby the central densities $\rhoh$ scatter more than the averaged $\rhoavdm$.
For two reasons, $\rhoavdm$ quantifies the actual dark matter density more robustly than
$\rhoh$. Firstly, our estimate of the very central dark matter density
depends strongly on the assumed halo profile. Instead, averaged over $2 \, \reff$ differences
between logarithmic halo fits and fits with NFW-profiles are small compared
to the statistical errors (averaged over
the Coma sample NFW-fits yield $0.1 - 0.2$ dex higher $\rhoavdm$ than fits with logarithmic
halos).

Secondly, the most significant differences between $\rhoh$ and $\rhoavdm$ occur in the
four Coma galaxies with distinct halos discussed
in Sec.~\ref{sec:scaling}. These galaxies have young central stellar populations. 
If the bulk of stars in these galaxies is old, however, then the related radial increase of 
the stellar mass-to-light ratio could contribute to their small $\rh$ and large $\rhoh$. 
This, because in our models it is assumed that the stellar
mass-to-light ratio is radially constant. By construction then, any increase in the 
mass-to-light ratio with radius (being either due to a stellar population gradient or 
due to dark matter) is attributed to the halo component. In galaxies with a significant
increase of the stellar $M/L$ with radius, 
the 'halo' component of the model thereby has to account for both, additional stellar and 
possible dark mass \citep{Tho06}. Any contamination with 
stellar mass will be largest at small radii, where the increase in the stellar $M/L$ dominates
the shape of the mass profile. The averaging radius in equation (\ref{avrhodef}) is 
therefore chosen as large as possible. The
value of $2 \, \reff$ is a compromise for the whole sample, because the kinematic
data extend to $1-3 \, \reff$ and the averaging should not go much beyond the last data point.

Compared to the Coma galaxies, the majority of \ger galaxies have larger $\rhoh$. After
averaging inside $2 \, \reff$, the halo densities in both samples become comparable, however.
In this respect, the \ger galaxies again resemble the four Coma galaxies with
young stellar cores.

Dotted and dashed lines in Fig.~\ref{fig:rho_l} show spiral galaxies. 
Their halo densities need not to be averaged before comparison, because 
core sizes of spirals (in the considered luminosity interval) are larger than 
$2 \, \reff$.
Averaged over the whole sample, we find dark matter densities in Coma early-types 
a factor of $6.8$ higher than in spirals of the same luminosity. If early-types
are compared to spirals of the same stellar mass, then the overdensity amounts to a factor of $13.5$.
Does this imply that spirals and ellipticals 
of the same luminosity have formed in different dark matter halos?

\section{Baryonic contraction}
\label{sec:contraction}
Even if ellipticals and spirals would have formed in similar
halos, then the final dark matter densities after the actual galaxy formation process could be different,
since the baryons in ellipticals and spirals are not distributed in the same way.
This effect can be approximated as follows: assume that (a spherical) 
baryonic mass distribution $\mstar(r)$ condenses slowly out of an original halo+baryon
distribution $M_i(r)$. The halo responds adiabatically and contracts into the 
mass distribution $M_\mathrm{DM}(r)$. If the original particles move on circular orbits then
\begin{equation}
\label{adiaeq}
r \left[ \mstar(r) + M_\mathrm{DM}(r) \right] = r_i M_i (r_i)
\end{equation}
turns out to be an adiabatic invariant \citep{Blu86}. 

In case of the Coma galaxies, $\mstar$ and $M_\mathrm{DM}$ are known from the dynamical
modeling and equation (\ref{adiaeq}) can be solved for $M_i$\footnote{Because of the large core-radii in some galaxies 
(cf. Fig.~\ref{fig:rhvh}c,d) it is not always possible to find $M_i$ for logarithmic halos.
Therefore, here we only consider the best-fit NFW-halo of each galaxy.}. It characterizes the
original halo mass distribution before the actual galaxy formation. 
Had a disk with baryonic mass $\mstar\disk$ grown in this original halo -- instead of an 
early-type -- then the halo contraction would have been different such that in general 
$M_\mathrm{DM}\disk \ne M_\mathrm{DM}$. The difference between $M_\mathrm{DM}\disk$ and $M_\mathrm{DM}$ actually
determines how much dark matter densities of ellipticals and spirals would
differ, if both had formed in the same original halos. To quantify this further, let's consider the spherically averaged mass distribution of a thin exponential 
disk for $\mstar\disk$ \citep[e.g.][]{Blu86}. It is fully determined by a scale-radius and 
a mass. For a given Coma elliptical with luminosity $L_B$, the scale-radius and the mass of a 
realistic disk with the same luminosity can be taken from
equations (\ref{spireff}) and (\ref{eq:mlspir}). Then, given $\mstar\disk$ and 
$M_i$ (the recontracted Coma galaxy halo), equation (\ref{adiaeq}) can be solved 
for the baryon-contracted halo $M_\mathrm{DM}\disk$ around the comparison disk 
(cf. \citealt{Blu86}). Once $M_\mathrm{DM}\disk$ is known, the average $\rhoavdm\disk$ 
follows directly.

\begin{figure*}\centering
\begin{minipage}{140mm}
\centering
\includegraphics[width=135mm,angle=0]{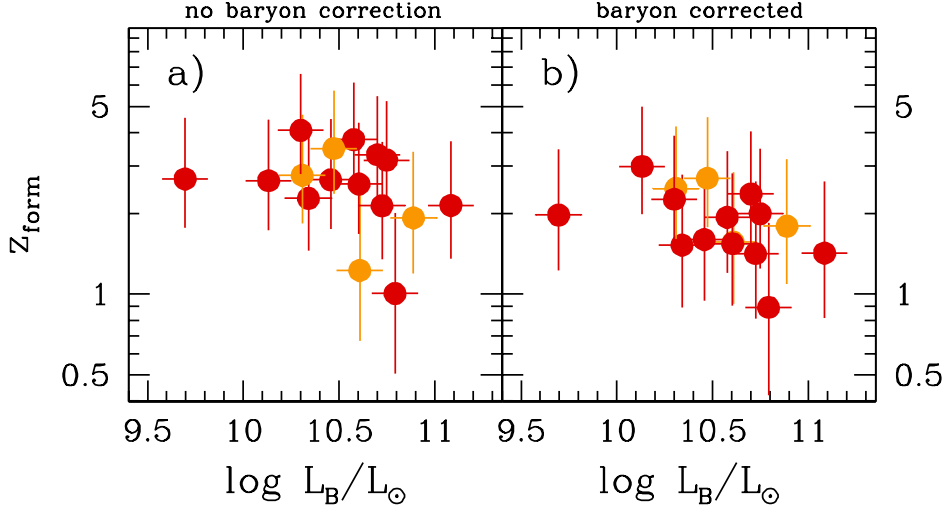}
\caption[]
{Estimated formation redshifts $\zform$ of Coma galaxy halos versus luminosity
without baryon correction (a) and with baryon correction (b). 
Symbols and colors as in Fig.~\ref{fig:rhvh}.}
\label{fig:zl}
\end{minipage}
\end{figure*}

If ellipticals and spirals (of the same
$L_B$) would have formed in the same halos, then
\begin{equation}
\label{dbar}
\dbar \equiv \frac{\rhoavdm}{\rhoavdm\disk}
\end{equation}
should fully account for the observed ratio of elliptical to spiral dark matter densities. 
However, averaged over the Coma sample we find $\dbar \approx 2$ and, thus, that
the higher baryon concentration in early-types is not sufficient to explain the factor of 
$6.8$ between the dark matter densities of ellipticals and spirals at
constant luminosity.

In general, the observed dark matter density
ratio $\dobs$ between ellipticals and spirals will be a combination of a difference in the 
halo densities before baryon infall and a factor that comes from the baryons.
Let $\dhalo$ denote the baryon-corrected dark matter density ratio, then the 
simplest assumption is
\begin{equation}
\label{eq:bcor}
\dobs = \dbar \times \dhalo,
\end{equation}
with $\dbar$ from equation (\ref{dbar}). After applying this approximate baryon
correction, dark matter densities in Coma ellipticals are still a factor of 
$\dhalo = 3.4$ higher than in spirals of the same luminosity. If the comparison 
is made at the same stellar mass, then $\dhalo = 6.4$. 
Note that our baryonic contraction corrections are likely upper limits, because in equation
(\ref{eq:mlspir}) we only account for the stellar mass in the disk. In the presence of
gas, the baryonic disk mass will be larger and so will be the halo contraction. The 
dark matter density contrast relative to the original elliptical will be therefore smaller.

Concluding, the differences between the baryon distributions of ellipticals and spirals are
not sufficient to explain the overdensity of dark matter in ellipticals relative
to spirals of the same luminosity or stellar mass. Ellipticals and spirals 
have not formed in the same halos. Instead, the higher dark matter density in ellipticals 
points to an earlier assembly redshift.

\section{The dark-halo assembly epoch of Coma early-types}
\label{sec:assembly}
In order to evaluate the difference between elliptical and spiral galaxy assembly
redshifts quantitatively,
let's assume that dark matter densities scale with the
mean density of the universe at the assembly epoch, i.e. $\rhodm \propto (1+\zform)^3$
(we will discuss this assumption in Sec.~\ref{subsec:sams_zzz}).
Let $\zform\el$ and $\zform\spir$ denote the formation redshifts of ellipticals
and spirals, respectively, then
\begin{equation}
\label{zscale}
\frac{1+\zform\el(L_B)}{1+\zform\spir(L_B)} = \left( \frac{\rhodm\el(L_B)}{\rhodm\spir(L_B)} \right)^{1/3}
\end{equation}
\citep{G01}, with $\rhodm$ some measure of the dark matter density, e.g. $\rhodm = \rhoavdm$. 
Equation (\ref{zscale}) can be solved for
\begin{equation}
\label{zformdef}
\zform\el = \left[ 1+\zform\spir \right] \times \delta^{1/3} - 1,
\end{equation}
where we have omitted the dependency of dark matter densities 
and formation redshifts on $L_B$ and defined $\delta = \rhodm\el/\rhodm\spir$.
Equation (\ref{zformdef}) allows
to calculate formation redshifts of Coma ellipticals from $\zform\spir$ and the 
observed $\delta$. Two estimates based on different assumptions about
$\delta$ and $\zform\spir$ are shown in Fig.~\ref{fig:zl} and are further discussed
below. For each case, formation redshifts were calculated with both
disk halo scaling laws shown in Fig.~\ref{fig:rho_l} and the two results were averaged.

\begin{figure*}\centering
\begin{minipage}{140mm}
\centering
\includegraphics[width=135mm,angle=0]{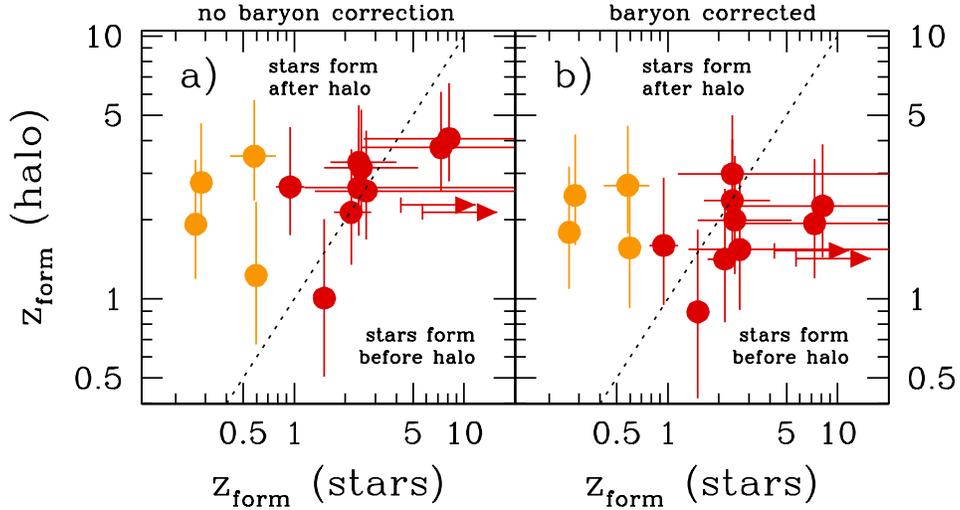}
\caption[]
{As Fig.~\ref{fig:zl}, but estimated formation redshifts $\zform$ of Coma galaxy halos 
are plotted versus the formation epoch $z(\tau_0)$ of the central stellar population 
(from column 7 of Tab.~\ref{tab:gals2}); dotted lines: one-to-one relations. 
Symbols and colors as in Fig.~\ref{fig:rhvh}.}
\label{fig:zz}
\end{minipage}
\end{figure*}

Raw formation redshifts without any baryon correction ($\delta = \dobs$)
are shown in Fig.~\ref{fig:zl}a (and listed in column 3 of Tab.~\ref{tab:gals2}). We considered
a wide range of spiral galaxy formation redshifts $\zform\spir \in [0.5,2]$ and the
related uncertainty in $\zform\el$ is indicated by vertical bars. Our fiducial value
is $\zform\spir \equiv 1$, because regular disks become rare beyond $z \ga 1$ 
\citep{Con05}. We did not allow for a luminosity dependence of spiral galaxy
formation times. For the Coma ellipticals we then find $\zform\el$ ranging from
$\zform\el \approx 0.5$ to $\zform\el \approx 5$, with the majority of galaxies having
formed around $\zform\el \approx 3$. Brighter galaxies have assembled later than fainter 
galaxies.

Coma galaxy assembly redshifts shown in Fig.~\ref{fig:zl}b (see also column 5 
of Tab.~\ref{tab:gals2}) include the baryon correction
of Sec.~\ref{sec:contraction}, because we used $\delta=\dhalo$ in equation (\ref{zformdef}). 
Moreover, because fainter spirals have denser halos than brighter ones,
we allowed for a luminosity dependent $\zform\spir(L)$. In analogy to
equation (\ref{zscale}) assume
\begin{equation}
\label{zfspir}
\frac{1+\zform\spir(L)}{1+\zform\spir(L_0)} = \left( \frac{\rhodm\spir(L)}{\rhodm\spir(L_0)} \right)^{1/3}
\end{equation}
and $z\spir_0 = 1$ for a reference luminosity $\log L_0/L_\odot = 10.5$ (the dashed 
line in Fig.~\ref{fig:z_sam}c illustrates the resulting $\zform\spir$ as a function of $L$).
Because $\dhalo \la \dobs$, the baryon corrected $\zform\el$ in
Fig.~\ref{fig:zl}b are lower than the 
uncorrected ones in Fig.~\ref{fig:zl}a. The typical assembly redshift reduces to 
$\zform\el \approx 2$, as compared to $\zform\el \approx 3$ without the correction. 
The baryon correction is mostly smaller than the uncertainty related to our ignorance
about $\zform\spir$ (vertical bars in Fig.~\ref{fig:zl}b correspond to 
$z\spir_0 \in [0.5,2]$). The trend for lower $\zform\el$ in brighter
galaxies is slightly diminished by the baryon correction such that the
dependency of $\zform\el$ on $L$ in Fig.~\ref{fig:zl}b mainly reflects 
the luminosity dependence of spiral galaxy assembly redshifts.

Fig.~\ref{fig:zz} compares halo assembly redshifts
with central stellar population ages $\tau_0$ from \citet{Meh03}. (We use 
$H_0 = 70 \, \kms \Mpc^{-1}$, $\Omega_\Lambda = 0.75$ and $\Omega_m = 0.25$ to transform
ages into redshifts.)
Largely independent from applying the baryon correction or not, the
agreement between the two redshifts is fairly good for about half of 
our sample. Among the remaining galaxies, some have halos which appear younger 
than their central stellar
populations. This could indicate that the stellar ages are 
overestimated (they are sometimes larger than the age of the universe in the adopted 
cosmology; cf. Tab.~\ref{tab:gals2}). It could also point at
these galaxies having grown by dry merging. In a dry merger, the dark matter
density can drop, but the stellar ages stay constant.
In Coma galaxies with young stellar cores, the 
halo assembly redshifts are instead larger
than the central stellar ages. This indicates some secondary star-formation
after the main epoch of halo assembly.

\section{Comparison with semi-analytic galaxy formation models}
\label{sec:semi-analytic}
In the following we will compare our results to semi-analytic galaxy 
formation models. To this end we have constructed a comparison sample of synthetic
ellipticals and spirals using the models of \citet{deL07}, which 
are based on the Millennium simulation \citep{Spr05}. 
Comparison ellipticals are selected to rest in dark matter cluster structures 
with virial masses larger than $M_\mathrm{vir} > 10^{15} M_\odot$ and to obey 
$M_{B,\mathrm{bulge}}-M_B<0.4$ \citep{Sim86}. We ignore galaxies at the centers
of simulated clusters since we have omitted the two central Coma
galaxies from the analysis in this paper. Likewise, we exclude 
from the comparison galaxies that have been stripped-off their entire halo, because
the only Coma galaxy that possibly lacks dark matter inside
$3\, \reff$ has been excluded from the analysis in this paper as well 
(cf. Sec.~\ref{sec:data}). Isolated field spirals are drawn from
objects with $M_{B,\mathrm{bulge}}-M_B>1.56$ in the semi-analytic models \citep{Sim86}.

Simulated galaxies were chosen randomly
from the catalogue of \citet{deL07} in a way such that each of six luminosity intervals 
(between $M_B = -17$ and $M_B = -23$; width $\Delta M_B = 1.0$) contains
roughly 50 galaxies. We use dust-corrected luminosities $M_B$ of the semi-analytic
models.

\subsection{Dark matter density}
Dark matter halos of simulated galaxies are reconstructed from
tabulated virial velocities $v_\mathrm{vir}$, virial radii $r_\mathrm{vir}$,
and maximum circular velocities
$v_\mathrm{max}$ as follows. It is assumed that the halos can be approximated by an
NFW-profile (cf. equation \ref{nfwdef}), in which case the circular velocity profile reads
\begin{equation}
\label{nfwcirc}
\left( \frac{v_\mathrm{circ}(r)}{v_\mathrm{vir}} \right)^2 = 
\frac{1}{x} \frac{\ln(1+cx)-cx/(1+cx)}{\ln(1+c) - c/(1+c)}.
\end{equation}
Here $x=r/r_\mathrm{vir}$ and the halo concentration is defined by $c=r_\mathrm{vir}/r_s$. 
The maximum circular velocity $v_\mathrm{max}$ of an NFW halo 
occurs at $r \approx 2 r_\mathrm{vir}/c$ \citep{Nav96}, such that (with
equation \ref{nfwcirc})
\begin{equation}
4.63 \left( \frac{v_\mathrm{max}}{v_\mathrm{vir}} \right)^2 = 
\frac{c}{\ln(1+c)-c/(1+c)}.
\end{equation}
Using the tabulated $v_\mathrm{vir}$ and $v_\mathrm{max}$ this equation can be numerically
solved for the halo concentration $c$, which in turn determines $r_s = r_\mathrm{vir}/c$
and, thus, the entire NFW profile of the halo. 

Before compared to the Coma galaxy models, halo densities are averaged 
within $2 \, \reff$ (cf. equation \ref{avrhodef}).
In case of simulated spirals we use effective radii from the empirical 
relation (\ref{spireff}). For ellipticals, we assume
\begin{equation}
\label{ellreff}
\left( \frac{\reff}{\kpc} \right) = 15.34 \left( \frac{L_B}{10^{11} \, \lsun} \right)^{1.02},
\end{equation}
which is a fit to the Coma data.

\begin{figure*}\centering
\begin{minipage}{140mm}
\centering
\includegraphics[width=135mm,angle=0]{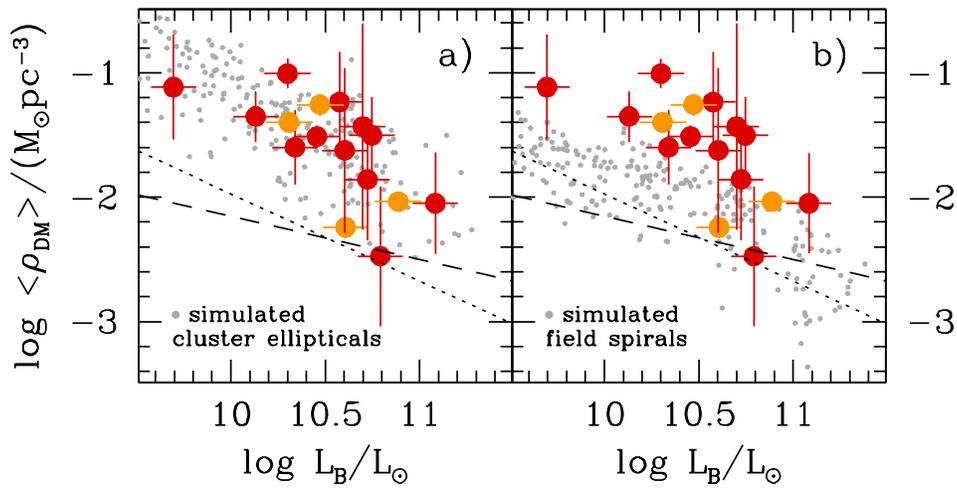}
\caption[]
{Average dark matter density $\rhoavdm$ versus $L_B$ in simulated 
cluster ellipticals (a) and in simulated field spirals (b). Large symbols and
lines as in Fig.~\ref{fig:rhvh}. Simulated galaxies from the semi-analytic models of 
\citet{deL07}.}
\label{fig:dm_sam}
\end{minipage}
\end{figure*}

Fig.~\ref{fig:dm_sam} shows that the average dark matter densities $\rhoavdm$ of the Coma
early-types match fairly well with semi-analytical models. 
This is remarkable, because the simulations do not take into account the
halo response during baryon infall. Therefore, either the net effect of the baryons 
on the dark matter distribution is small in the analyzed population of galaxies or
there is actually a mismatch between the halos of observed galaxies 
and the $N$-body models. It may also be that real galaxies do not have
maximum stellar masses. This can be checked by the comparison of dynamically derived
stellar mass-to-light ratios with independent stellar population 
synthesis models (Thomas et al., in preparation).

Similarly to what is found in real galaxies, the dark matter densities 
of spirals are lower than in ellipticals in semi-analytic models 
(cf. Fig.~\ref{fig:dm_sam}), but the density contrast in observed galaxies is larger. 
Again, a major uncertainty here is that the simulations
do not take into account the gravitational effect of the baryons.

\subsection{Assembly redshift}
\label{subsec:sams_zzz}
\begin{figure*}\centering
\begin{minipage}{160mm}
\centering
\includegraphics[width=155mm,angle=0]{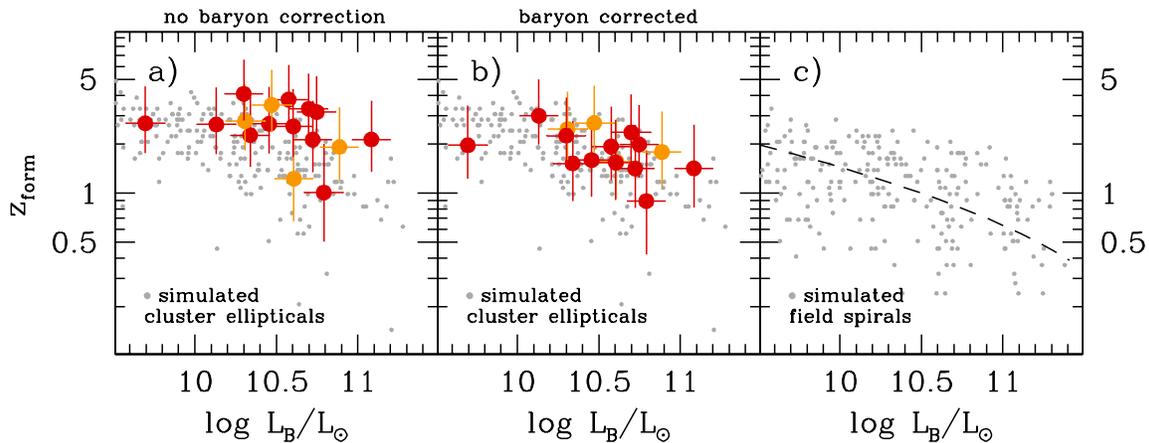}
\caption[]
{Comparison between assembly redshifts $\zform$ of simulated and
observed galaxies (symbols as in Fig.~\ref{fig:dm_sam}). Both formation redshift
estimates for Coma galaxies with and without baryon correction are shown (panels a,b).
For comparison, panel (c) shows formation redshifts of simulated spirals (dots) and of
observed spirals (dashed-line; cf. equation \ref{zfspir}). 
Simulated galaxies from the semi-analytic models of \citet{deL07}.}
\label{fig:z_sam}
\end{minipage}
\end{figure*}

Formation redshifts of simulated and observed galaxies are compared in 
Fig.~\ref{fig:z_sam}. Coma
galaxy $\zform$ are from Sec.~\ref{sec:assembly} and both
cases discussed there -- with and without baryon correction -- are shown separately
in panels (a) and (b), respectively. Formation redshifts of simulated galaxies 
are defined as the earliest redshift, when a halo has 
assembled 50 percent of its mass. Since we are mainly interested in cluster 
ellipticals, we need to take into account that interactions between the cluster halo and
a galaxy's subhalo cause a mass-loss in the latter. Although cluster-galaxy interactions
happen in both simulated and observed galaxies, the mass-loss in the simulations may
be overestimated because of the finite numerical resolution and the neglect of the
baryon potential. In particular, for simulated subhalos with very low masses at $z=0$ 
the derived formation redshifts may be artificially high, when defined according to the
assembly of half of the final mass. To avoid such artificially large assembly redshifts,
we define $\zform$ of simulated galaxies as the earliest time when half of the
maximum mass was assembled, that a single progenitor in the merger tree of given galaxy 
had at some redshift. Our assumption is that even if dynamical interactions between cluster
and galaxy halos take place, they do not significantly affect the very inner regions $<2 \, \reff$ of 
interest here. In case of field spirals, formation redshifts defined either from the
final or from the maximum mass are very similar.
 
Without a baryon correction, our estimates of Coma galaxy formation redshifts are on
average higher than in the semi-analytic models (Fig.~\ref{fig:z_sam}a).
This,
although (1) the dark matter densities of ellipticals match with the simulations and
(2) our assumption about the formation redshifts of spirals ($\zform\spir \approx 1$) is
consistent with the simulations. The origin for the offset between Coma galaxies and
semi-analytic models in 
Fig.~\ref{fig:z_sam}a is that the density contrast between halos of ellipticals and spirals
is larger in observed galaxies than in the simulations.
After applying the baryon correction, the Coma galaxy
formation redshifts become consistent with the simulations (Fig.~\ref{fig:z_sam}b). 
This result indicates that the discrepancy between the measured
and the simulated density ratio $\rhodm\el/\rhodm\spir$ is due to baryon effects. 

\begin{figure*}\centering
\begin{minipage}{140mm}
\centering
\includegraphics[width=135mm,angle=0]{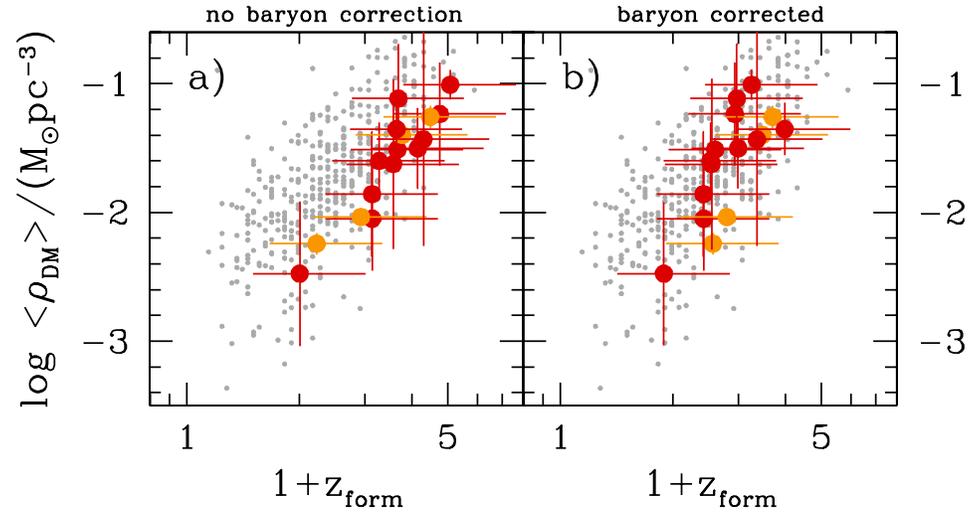}
\caption[]
{Average halo density $\rhoavdm$ versus assembly redshift. Large symbols:
Coma galaxies without baryon correction (a) and with
baryon correction (b). Small symbols: simulated cluster ellipticals
and simulated field spirals from the semi-analytic models of \citet{deL07}.}
\label{fig:rho_zf_sam}
\end{minipage}
\end{figure*}

Our Coma galaxy formation redshifts are based on the assumption that
$\rhoavdm \propto (1+\zform)^3$. Fig.~\ref{fig:rho_zf_sam} shows $\rhoavdm$ versus
$(1+\zform)$ explicitly. Independent of including a baryon correction or not, the 
slope of the relationship between $\rhoavdm$ and $(1+\zform)$ in the
Coma galaxies is roughly parallel to simulated $N$-body halos. This
confirms that our assumption for the scaling between $\rhoavdm$ and $\zform$ 
is approximately consistent with the cosmological simulations.

Concerning the absolute values of the dark matter densities it
has been already stated above that they are only consistent with the simulations 
if either the net effect of the baryons is
zero in the case of ellipticals or if galaxies do not have maximum stellar masses.
The former case would imply that halos of spiral galaxies experience a net expansion during 
the baryon infall (several processes have been proposed for this, e.g. \citealt{Bin01}). 

\citet{deL06} quote a stellar assembly redshift below $z<1$ for simulated ellipticals 
more massive than $M_\ast > 10^{11} \, \msun$. The halo assembly redshifts in 
Fig.~\ref{fig:z_sam} are mostly above $z>1$. In part, this is due to the fact that
we only consider semi-analytic galaxies in high-density environments similar to Coma. In addition, 
formation redshifts defined according to the stellar mass assembly and the halo assembly,
respectively, are not always equal. For example, in our comparison sample of simulated
cluster ellipticals we find
an average dark halo assembly redshift $\langle \zform \rangle = 1.50$ for galaxies 
more massive than $M_\ast > 10^{11} \, \msun$. Evaluating for the same galaxies 
the redshift $\zform^\ast$ (when half the stellar mass is assembled) yields 
$\langle\zform^\ast \rangle = 1.07$. 
That $\zform \ga \zform^\ast$ is plausible if some star formation is going on 
between $0 \le z \le \zform$ in the progenitor and/or in the subunits that are to be 
accreted after $\zform$. It should also be noted that the simulations do not take into account 
stellar mass-loss due to tidal interactions.

\section{Summary}
\label{sec:summary}
We have presented dark matter scaling relations derived from axisymmetric, orbit-based 
dynamical models of flattened and rotating as well as non-rotating 
Coma early-type galaxies. Dark matter halos in these galaxies follow similar
trends with luminosity as in spirals. Thereby, the majority of Coma
early-types -- those with old stellar populations -- have halo core-radii $\rh$ 
similar as in spirals with the same $B$-band 
luminosity, but their asymptotic halo velocities are about $2.4$ times higher. 
In contrast, four Coma early-types -- with young central stellar 
populations -- have halo velocities of the same order as in comparably bright spirals, but 
their core-radii are smaller by a factor of 4. 
Differences between spirals and ellipticals 
increase, when the comparison is made at the same stellar mass.
The average halo density inside $2 \, \reff$ 
exceeds that of comparably bright spirals by about a factor of $6.8$. 
If the higher baryon concentration in ellipticals is taken into account, the
excess density reduces to about a factor of 3, but if ellipticals and spirals
are compared at the same stellar mass, then it is again of the order of $6.5$.

Our measured dark matter densities match with a comparison sample of simulated
cluster ellipticals constructed from the semi-analytic galaxy formation models
of \citet{deL07}. These synthetic ellipticals have $\zform \approx 0.5-4$ and
higher dark matter densities than simulated field spirals, which
are appear on average around $\zform\spir \approx 1$.

Assuming for local spirals $\zform\spir = 1$ as well, and assuming further 
that the inner dark matter density
scales with the formation redshift like $(1+\zform)^3$, our results imply
that ellipticals have formed $\Delta \zform \approx 1-2$ earlier than spirals.
Without baryon correction, we find an average formation redshift around $\zform \approx 3$,
which is slightly larger than in semi-analytic galaxy formation models. Accounting
for the more concentrated baryons in ellipticals, the average formation redshift
drops to $\zform \approx 2$. 

For about half of our sample, dark halo formation redshifts match
with constraints derived from stellar populations \citep{Meh03}: the assembly epoch of 
these (old) early-types coincides with the epoch of formation of their stellar components.

\acknowledgements
We thank Ortwin Gerhard and the anonymous referee for comments and suggestions that
helped to improve the manuscript. JT acknowledges financial support by the 
Sonderforschungsbereich 375 'Astro-Teilchenphysik' of the Deutsche Forschungsgemeinschaft.
EMC receives support from grant CPDA068415/06 by Padua University.
The Millennium Simulation databases used in this paper and the web application providing 
online access to them were constructed as part of the activities of the German 
Astrophysical Virtual Observatory.

\appendix

\section{GMP3414 and GMP4822}
\label{app:3414}
\begin{figure*}\centering
\begin{minipage}{100mm}
\centering
\includegraphics[width=95mm,angle=0]{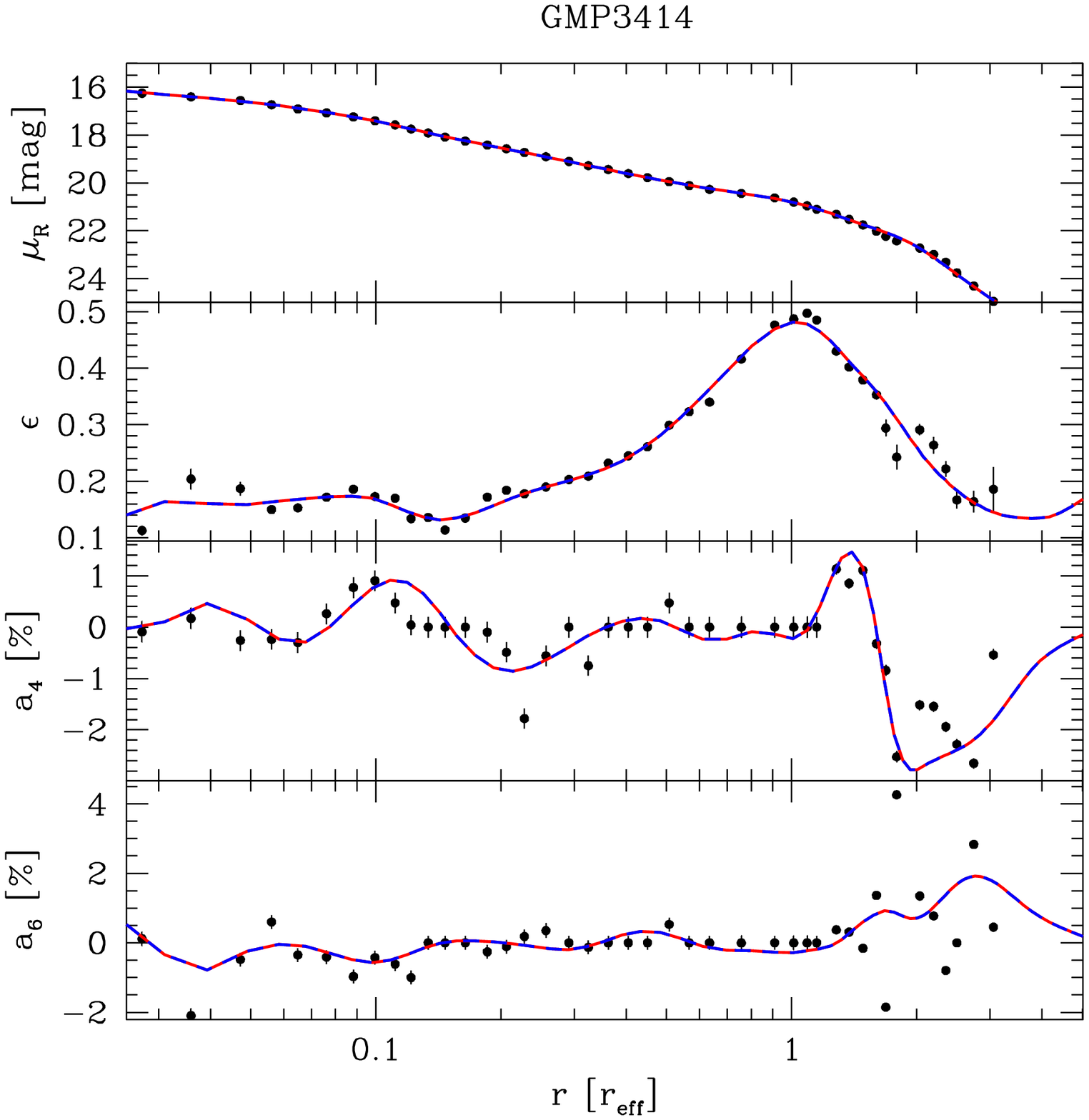}
\includegraphics[width=95mm,angle=0]{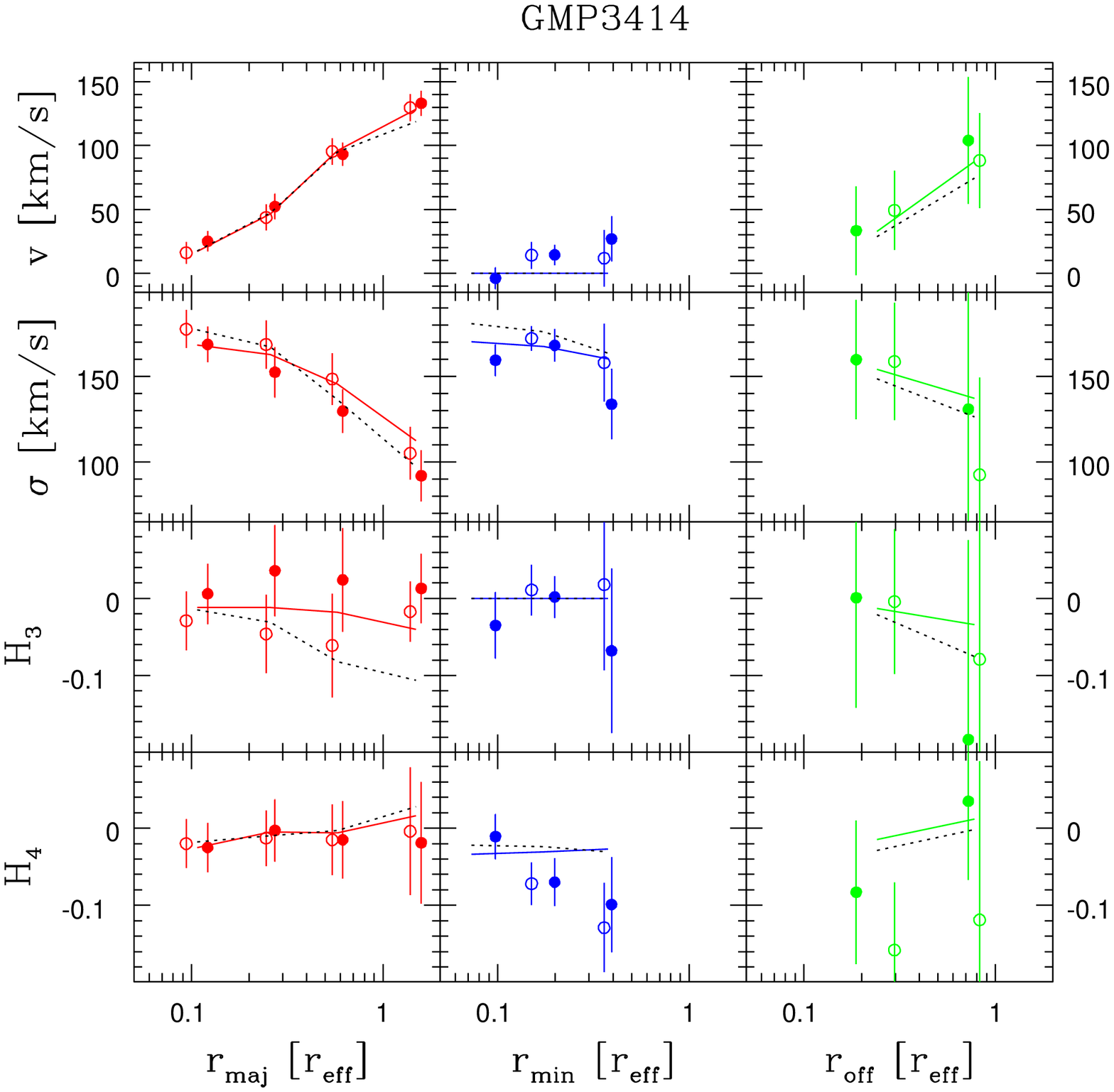}
\caption[]
{Upper panel: joint ground-based and HST photometry of GMP3414/NGC4871. Lines:
best-fit deprojection (red) and its edge-on reprojection (blue). Lower panel: 
stellar kinematics along major axis (left/red), along the minor axis (middle/blue)
and along a third axis parallel to the major axis with an offset
of $\reff/2$ (right/green); filled and open circles refer to different sides of the slits; 
dotted: best-fit model without dark matter.}
\label{fig:3414}
\end{minipage}
\end{figure*}

\begin{figure*}\centering
\begin{minipage}{100mm}
\centering
\includegraphics[width=95mm,angle=0]{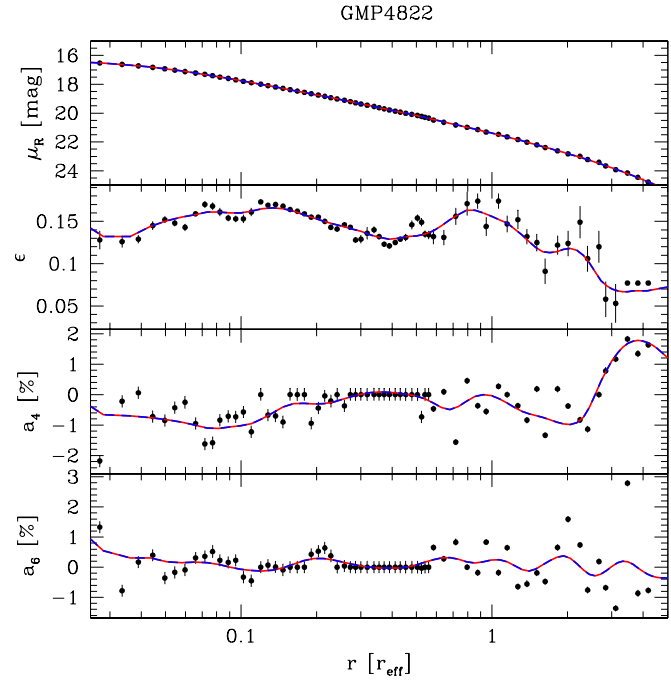}
\includegraphics[width=95mm,angle=0]{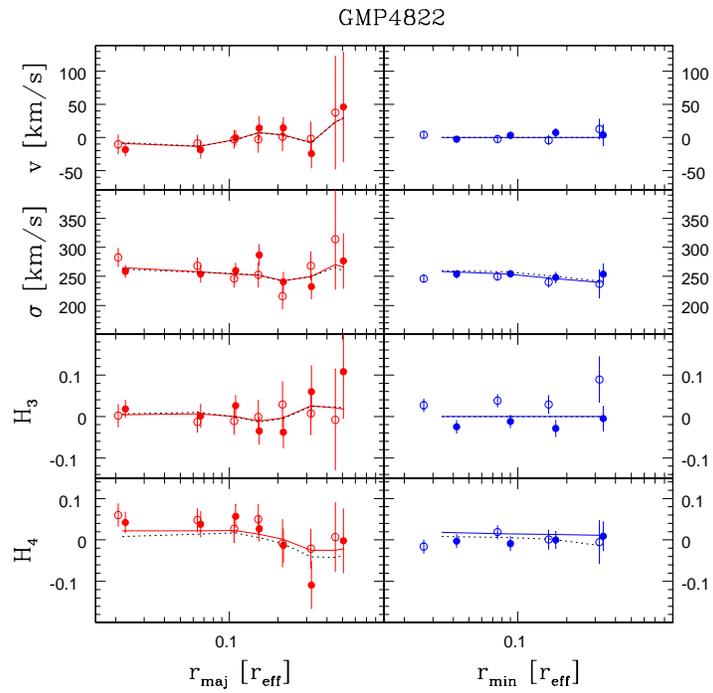}
\caption[]
{As Fig.~\ref{fig:3414} but for GMP4822/NGC4841A.}
\label{fig:4822}
\end{minipage}
\end{figure*}

The best-fit model parameters for the galaxies GMP3414 and GMP4822 
(which were not included in the
original sample of \citealt{Tho07}) are given in Tab.~\ref{tab:3414}. The table is
similar to Tab.~2 of \citet{Tho07} and we refer the reader to this paper, in case more
detailed information about the parameter definitions are required. The best-fit models
with and without dark matter halo are compared to the observations in 
Figs.~\ref{fig:3414} and \ref{fig:4822}. In both galaxies the
best-fit inclination is $i=90\degr$, but the 68 percent confidence regions include 
models at $i\ge70\degr$ (GMP3414) and $i\ge50\degr$ (GMP4822).

GMP3414 and GMP4822 were observed with the Wide Field Planetary Camera 2 (WFPC2) on board
the {\it HST} as part of the {\it HST} proposal 10844 (PI: G. Wegner). For each galaxy
two exposures
with 300s each were taken with the filter F622W. Four other objects were previously
observed as part of this proposal and a full description of the
respective observational parameters and the data analysis is given in 
\citet{Cor08}. A table with the final photometric parameters will be published
in the journal version of this paper.

\clearpage
\begin{deluxetable*}{lcccccccc}
\tablecaption{Dark matter scaling relations.\label{tab:fits}}
\tablewidth{120mm}
\tablehead{
\multicolumn{2}{c}{relation} &
\multicolumn{2}{c}{$\log y=a+b \log x$} &
 &
 &
 &
\\
\colhead{$y$} &
\colhead{$x$} &
\colhead{$a$} &
\colhead{$b$} &
\colhead{$\chi^2_\mathrm{red}$} & 
\colhead{rms} &
\colhead{$\langle \Delta \log y \rangle$}&
\colhead{$\cal P$}&
\colhead{figure}\\
\colhead{(1)}&
\colhead{(2)}&
\colhead{(3)}&
\colhead{(4)}&
\colhead{(5)}&
\colhead{(6)}&
\colhead{(7)}&
\colhead{(8)}&
\colhead{(9)}
}
\startdata
\multicolumn{8}{c}{fits to all galaxies}\\
\hline
$\frac{\rh}{\kpc}$             & $\frac{L_B}{10^{11} \, \lsun}$    & $1.24 \pm 0.14$ & $0.55 \pm 0.26$& 1.24 & 0.35 & 0.39 &   0.010& \\
$\frac{\rh}{\kpc}$             & $\frac{\mstar}{10^{11} \, \msun}$ & $0.71 \pm 0.12$ & $0.90 \pm 0.28$& 0.85 & 0.31 & 0.39 &   0.002& \\
$\frac{\vh}{\kms}$             & $\frac{L_B}{10^{11} \, \lsun}$    & $0.52 \pm 0.12$ & $0.07 \pm 0.23$& 1.62 & 0.19 & 0.21 &   0.109& \\
$\frac{\vh}{\kms}$             & $\frac{\mstar}{10^{11} \, \msun}$ & $0.33 \pm 0.07$ & $0.45 \pm 0.16$& 1.04 & 0.17 & 0.21 &   0.002& \\
$\frac{\rhoh}{\msun \pc^3}$    & $\frac{\rh}{\kpc}$               & $0.67 \pm 0.64$ & $-1.99 \pm 0.57$& 0.41 & 0.41 & 0.66 &   0.001& \\
$\frac{\rh}{\kpc}$             & $\frac{\reff}{\kpc}$             & $0.48 \pm 0.23$ & $0.79 \pm 0.32$& 1.15 & 0.31 & 0.39 &    0.026& \\
$\frac{\rhoh}{\msun \pc^3}$    & $\frac{L_B}{10^{11} \, \lsun}$    & $-1.87 \pm 0.18$ & $-1.28 \pm 0.33$& 0.75 & 0.47 & 0.66 & 0.019& \ref{fig:rho_l}a \\
$\frac{\rhoh}{\msun \pc^3}$    & $\frac{\mstar}{10^{11} \, \msun}$ & $-0.77 \pm 0.19$ & $-1.57 \pm 0.38$& 0.94 & 0.52 & 0.66 & 0.058& \ref{fig:rho_l}b \\
$\frac{\rhoavdm}{\msun \pc^3}$ & $\frac{L_B}{10^{11} \, \lsun}$    & $-2.36 \pm 0.14$ & $-1.56 \pm 0.24$& 1.31 & 0.38 & 0.32 & 0.004& \ref{fig:rho_l}c \\
$\frac{\rhoavdm}{\msun \pc^3}$ & $\frac{\mstar}{10^{11} \, \msun}$ & $-1.10 \pm 0.11$ & $-1.57 \pm 0.24$& 1.79 & 0.44 & 0.32 & 0.025& \ref{fig:rho_l}d \\
\hline\\
\multicolumn{8}{c}{fits omitting galaxies with young stellar cores}\\
\hline
$\frac{\rh}{\kpc}$             & $\frac{L_B}{10^{11} \, \lsun}$    & $1.54 \pm 0.21$ & $0.63 \pm 0.33$& 0.16 & 0.16 & 0.44 &    0.002&\ref{fig:rhvh}a\\
$\frac{\rh}{\kpc}$             & $\frac{\mstar}{10^{11} \, \msun}$    & $0.98 \pm 0.14$ & $0.54 \pm 0.29$& 0.18 & 0.17 & 0.44 & 0.007&\ref{fig:rhvh}b\\
$\frac{\vh}{\kms}$             & $\frac{L_B}{10^{11} \, \lsun}$    & $0.78 \pm 0.11$ & $0.21 \pm 0.21$& 0.14 & 0.07 & 0.24 &    0.016&\ref{fig:rhvh}c\\
$\frac{\vh}{\kms}$             & $\frac{\mstar}{10^{11} \, \msun}$    & $0.59 \pm 0.10$ & $0.19 \pm 0.17$& 0.12 & 0.07 & 0.24 & 0.008&\ref{fig:rhvh}d\\
$\frac{\rhoh}{\msun \pc^3}$    & $\frac{\rh}{\kpc}$               & $0.68 \pm 1.35$ & $-1.64 \pm 0.96$& 0.04 & 0.18 & 0.75  &   0.006&\ref{fig:rhoh_rh}\\
$\frac{\rh}{\kpc}$             & $\frac{\reff}{\kpc}$             & $0.79 \pm 0.21$ & $0.62 \pm 0.30$& 0.10 & 0.14 & 0.44 &     0.017&\ref{fig:rh_reff}\\
$\frac{\rhoh}{\msun \pc^3}$    & $\frac{L_B}{10^{11} \, \lsun}$    & $-1.81 \pm 0.36$ & $-1.02 \pm 0.54$& 0.31 & 0.34 & 0.75 &  0.028&\\
$\frac{\rhoh}{\msun \pc^3}$    & $\frac{\mstar}{10^{11} \, \msun}$ & $-0.94 \pm 0.17$ & $-0.81 \pm 0.43$& 0.31 & 0.36 & 0.75 &  0.090&\\
$\frac{\rhoavdm}{\msun \pc^3}$ & $\frac{L_B}{10^{11} \, \lsun}$    & $-2.01 \pm 0.20$ & $-1.04 \pm 0.30$& 0.77 & 0.31 & 0.41 &  0.012&\\
$\frac{\rhoavdm}{\msun \pc^3}$ & $\frac{\mstar}{10^{11} \, \msun}$ & $-1.12 \pm 0.11$ & $-0.74 \pm 0.25$& 0.86 & 0.33 & 0.41 &  0.020&\\
\enddata
\tablecomments{(1,2) Fitted quantities; (3,4) parameters of linear fit with errors;
(5) reduced $\chi^2_\mathrm{red}$ of the fit;
(6) rms-scatter in $\log y$; (7) mean error $\langle \Delta \log y \rangle$; 
(8) significance of the relation (probability $\cal P$ that there is no relation according to
a Spearman rank order correlation test);
(9) figure in which
the relation is shown.}
\end{deluxetable*}

\clearpage
\begin{deluxetable*}{lcccccc}
\tablecolumns{7}
\tablewidth{120mm}
\tabletypesize{\small}
\tablewidth{0pt}
\tablecaption{Halo densities, halo assembly redshfits and stellar ages\label{tab:gals2}}
\tablehead{
\colhead{galaxy} &
\colhead{$\log \frac{\rhoavdm}{\msun \pc^{-3}}$}&
\colhead{$\zform$}&
\colhead{$\log \frac{\rhoavdm/\dbar}{\msun \pc^{-3}}$}&
\colhead{$\zform$}&
\colhead{$\frac{\tau_0}{\Gyr}$}&
\colhead{$\zssp$}\\
\colhead{(1)} &
\colhead{(2)}&
\colhead{(3)}&
\colhead{(4)}&
\colhead{(5)}&
\colhead{(6)}&
\colhead{(7)}
}
\startdata
0144 & $-2.24 \pm 0.08$ & $1.23^{+1.11}_{-0.56}$ &  $-2.00 \pm 0.08$ & $1.56^{+1.56}_{-0.64} $ & $5.8  \pm 0.5$ &  $0.60^{+0.08}_{-0.07}$\\
0282 & $-1.51 \pm 0.09$ & $2.66^{+1.83}_{-0.92}$ &  $-1.98 \pm 0.09$ & $1.60^{+1.30}_{-0.65} $ & $7.7  \pm 0.8$ &  $0.95^{+0.20}_{-0.17}$\\
0756 & $-2.04 \pm 0.04$ & $1.92^{+1.46}_{-0.73}$ &  $-1.89 \pm 0.04$ & $1.79^{+1.40}_{-0.70} $ & $3.1  \pm 0.2$ &  $0.26^{+0.02}_{-0.02}$\\
1176 & $-1.40 \pm 0.07$ & $2.77^{+1.89}_{-0.94}$ &  $-1.61 \pm 0.07$ & $2.47^{+1.74}_{-0.87} $ & $3.3  \pm 0.4$ &  $0.28^{+0.04}_{-0.04}$\\
1750 & $-1.50 \pm 0.31$ & $3.15^{+2.08}_{-1.04}$ &  $-1.80 \pm 0.31$ & $1.99^{+1.50}_{-0.75} $ & $11.3 \pm 1.7$ &  $2.48^{+2.90}_{-0.98}$\\
2417 & $-1.62 \pm 0.66$ & $2.57^{+1.79}_{-0.89}$ &  $-2.01 \pm 0.66$ & $1.54^{+1.27}_{-0.64} $ & $11.5 \pm 2.4$ &  $2.65^{+\infty}_{-1.33}$\\
2440 & $-1.01 \pm 0.11$ & $4.07^{+2.54}_{-1.27}$ &  $-1.69 \pm 0.11$ & $2.25^{+1.63}_{-0.81} $ & $13.5 \pm 2.1$ &  $8.17^{+\infty}_{-5.61}$\\
3414 & $-1.35 \pm 0.20$ & $2.64^{+1.82}_{-0.91}$ &  $-1.60 \pm 0.20$ & $2.50^{+1.75}_{-0.88} $ & $11.2 \pm 2.7$ &  $2.40^{+\infty}_{-1.25}$\\
3510 & $-1.60 \pm 0.30$ & $2.27^{+1.64}_{-0.82}$ &  $-2.02 \pm 0.30$ & $1.52^{+1.26}_{-0.63} $ & $14.2 \pm 1.6$ &  $>4.26$\\
3792 & $-1.24 \pm 0.40$ & $3.76^{+2.38}_{-1.19}$ &  $-1.83 \pm 0.40$ & $1.93^{+1.47}_{-0.73} $ & $13.4 \pm 2.1$ &  $7.33^{+\infty}_{-4.84}$\\
3958 & $-1.11 \pm 0.42$ & $2.69^{+1.84}_{-0.92}$ &  $-1.82 \pm 0.42$ & $1.97^{+1.48}_{-0.74} $ & \nodata  & \nodata\\
4822 & $-1.43 \pm 0.83$ & $3.30^{+2.15}_{-1.08}$ &  $-1.65 \pm 0.83$ & $2.40^{+1.61}_{-0.77} $ & $11.2 \pm 1.3$ &  $2.40^{+1.61}_{-0.77}$\\
4928 & $-2.05 \pm 0.40$ & $2.14^{+1.57}_{-0.78}$ &  $-2.08 \pm 0.40$ & $1.42^{+1.21}_{-0.61} $ & $14.5 \pm 1.4$ &  $>5.70$\\
5279 & $-1.86 \pm 0.49$ & $2.13^{+1.57}_{-0.78}$ &  $-2.08 \pm 0.49$ & $1.42^{+1.21}_{-0.60} $ & $10.9 \pm 0.8$ &  $2.16^{+0.67}_{-0.45}$\\
5568 & $-2.47 \pm 0.56$ & $1.01^{+1.00}_{-0.50}$ &  $-2.40 \pm 0.56$ & $0.89^{+0.95}_{-0.47} $ & $9.6  \pm 0.6$ &  $1.50^{+0.26}_{-0.21}$\\
5975 & $-1.26 \pm 0.09$ & $3.49^{+2.24}_{-1.12}$ &  $-1.52 \pm 0.09$ & $2.70^{+1.85}_{-0.93} $ & $5.7  \pm 1.2$ &  $0.58^{+0.20}_{-0.16}$\\
\enddata
\tablecomments{(1) Galaxy id from \citet{GMP}; (2) average dark matter density $\rhoavdm$ inside $2 \, \reff$; (3) 
halo assembly redshift $\zform$ according to equation (\ref{zformdef}) with
$\delta = \dobs$ and $\zform\spir = 1$; (4) as column (2), but including the baryonic correction
$\dbar$ defined in equation (\ref{dbar}) (we do not derive an error estimate for the baryonic
contraction, but use the same errors in columns (2) and (4), respectively);
(5) $\zform$ as in column (3), but with baryon corrected $\delta = \dhalo$ 
and $\zform\spir(L)$ from equation (\ref{zfspir}); (6) central stellar age $\tau_0$ 
from Tab.~B.1 of \citet{Meh03} (GMP3958 has no age estimate because of its very 
low H$\beta$); (7) formation redshift $\zssp$ of the stars derived from column (6). 
In some galaxies the stellar age or its upper limit exceed the age of the universe in the 
adopted cosmology. In such cases only a lower limit is given for $\zssp$ or 
the upper redshift error is set equal to $\infty$, respectively.}
\end{deluxetable*}

\begin{deluxetable}{lccccccc}
\tablecaption{Model parameters for GMP3414 and GMP4822.\label{tab:3414}}
\tablecolumns{7}
\tablehead{
\colhead{galaxy} &
\colhead{fit} &
\colhead{$\Upsilon$}&
\colhead{$\rh$}&
\colhead{$\vh$}&
\colhead{$c$}&
\colhead{$q$}&
\colhead{$\chi^2$}\\
\colhead{(1)}&
\colhead{(2)}&
\colhead{(3)}&
\colhead{(4)}&
\colhead{(5)}&
\colhead{(6)}&
\colhead{(7)}&
\colhead{(8)}
}
\startdata
GMP3414 & SC & $6.0$ & & & & & $0.490$\\
& LOG & $4.5$ & $9.7$ & $356$ & & & $0.239$\\
& NFW & $4.0$ &  & & $15.30$ & $1.0$ & $0.238$\\
\hline
GMP4822 & SC & $6.5$ & & & & & $0.259$\\
& LOG & $5.5$ & $13.1$ & $552$ & & & $0.229$\\
& NFW & $5.0$ &  & & $6.71$ & $1.0$ & $0.232$
\enddata
\tablecomments{(1) galaxy id; (2) type of fit (SC: without dark matter;
LOG: logarithmic halo; NFW: halo profile from \citealt{Nav96});
(3) best-fit stellar $\Upsilon$ $[M_\odot/L_\odot]$ ($R_C$-band); (4,5) best-fit
logarithmic halo parameters $\rh$ $[\kpc]$ and $\vh$ $[\kms]$; (6,7) best-fit
NFW concentration $c$ and flattening $q$ (cf. \citealt{Tho07} for details); (8) achieved
goodness-of-fit $\chi^2$ (per data point).}
\end{deluxetable}

\end{document}